\documentclass[runningheads]{llncs}
\usepackage{graphicx} % Required for 
\graphicspath{ {./images/} }
\usepackage{url}
\usepackage[T1]{fontenc}
\usepackage[english]{babel}
\usepackage{subcaption}
\usepackage{caption}[singlelinecheck=false]
\usepackage{url}
\usepackage{colortbl}
\usepackage{amssymb}
\usepackage{stmaryrd}
\usepackage{amsmath}
\usepackage{mathrsfs}
\usepackage{amssymb}
\usepackage[left]{lineno}
\usepackage{xfrac}
\usepackage{booktabs}
\usepackage{nicefrac}
\usepackage[textsize=scriptsize,backgroundcolor=yellow!40]{todonotes}
\usepackage[nonumberlist,acronym,sanitize=none]{glossaries}
\glsdisablehyper
\usepackage{comment}
\usepackage[pdftex, colorlinks=true, hyperfootnotes=true, hyperindex=true,
            plainpages=false, pagebackref=false, pdfpagelabels=true, pdfstartview=FitH,
            linkcolor=purple, citecolor=teal, urlcolor=blue,
            bookmarks, bookmarksopen, bookmarksdepth=3]{hyperref}
\usepackage[capitalise,nameinlink]{cleveref}
\crefname{algocf}{Alg.}{Algs.}
\Crefname{algocf}{Algorithm}{Algorithms}
\crefname{section}{Sect.}{Sects.}
\Crefname{section}{Section}{Sections}
\usepackage{tkz-base}
\usetikzlibrary{decorations.pathmorphing,trees,snakes,arrows,shapes,automata,petri}
\usepackage{paralist}
\usepackage{multicol}
\usepackage{booktabs}
\usepackage[inline]{enumitem}
\usepackage{multirow}
\usepackage{rotating}
\usepackage{etaremune}
\usepackage[ruled,linesnumbered,vlined,algo2e]{algorithm2e}
\usepackage{marginnote}
\usepackage{mathtools}
\usepackage{mathabx}
\usepackage{adjustbox}
\usepackage{ifthen}
\usepackage[normalem]{ulem}
\usepackage{lineno}
\usepackage{soul}
\usepackage{floatrow}
\usepackage{listings}
\crefname{lstlisting}{Listing}{Listings}
\usepackage{lipsum}
\usepackage{makecell}
\usepackage{diagbox}
\usepackage[scientific-notation=false]{siunitx}
\usepackage{footmisc}
\usepackage{xspace}
\usepackage{ifdraft}
\usepackage{wrapfig}
\usepackage{orcidlink}

\usepackage{times}
\usepackage[subtle]{savetrees}%
\newacronym[\glslongpluralkey={Distributed Ledger Technologies}]{dlt}{DLT}{Distributed Ledger Technology}

\newacronym{ipfs}{IPFS}{InterPlanetary File System}
\newcommand{\IPFS}[0] {\Gls{ipfs}\xspace}

\newacronym{p2p}{P2P}{peer-to-peer}

\newacronym{abe}{ABE}{Attribute-Based Encryption}
\newcommand{\ABE}[0] {\Gls{abe}\xspace}

\newacronym{maabe}{MA-ABE}{Multi-Authority Attribute-Based Encryption}
\newcommand{\MAABE}[0] {\Gls{maabe}\xspace}

\newacronym{cpabe}{CP-ABE}{Ciphertext-Policy Attribute-Based Encryption}
\newcommand{\CPABE}[0] {\Gls{cpabe}\xspace}

\newacronym{ssl}{SSL}{Secure Sockets Layer}

\newacronym{dht}{DHT}{Distributed Hash Table}

\newacronym{mpc}{MPC}{Multi-party Computation}

\newglossaryentry{box}{ %
  name={Box},
  description={authority intialisation storage box},
  first={\glsentrydesc{box} (henceforth, \glsentrytext{Box} for short)},
  plural={Boxes},
  firstplural={\glsentrydesc{box}es (henceforth, \glsentryplural{box} for short)}
}

\newglossaryentry{rloc}{ %
	name={resource locator},
	description={a generic term to denote content-based links obtained via hashing (e.g., the IPFS link)},
}

\newacronym{cpmaabe}{CP-MA-ABE}{Ciphertext-Policy Multi-Authority Attribute-Based Encryption}

\newacronym{dk}{\textsl{dk}}{decryption key}
\newacronym{fdk}{\textsl{fdk}}{final decryption key}

\newacronym[\glslongpluralkey={Business Processes}]{bp}{BP}{Business Process}
\newacronym{bpi}{BPI}{Business Process Intelligence}
\newacronym{bpm}{BPM}{Business Process Management}
\newacronym{bpms}{BPMS}{Business Process Management System}
\newacronym{bpmn}{BPMN}{Business Process Model and Notation}
\newacronym{cpn}{CPN}{colored Petri net}
\newacronym{kpi}{KPI}{Key Performance Indicator}
\newacronym{ocbc}{OCBC}{Object-centric Behavioral Constraints}
\newacronym{soa}{SOA}{Service-Oriented Architecture}
\newacronym{pn}{PN}{Petri net}
\newacronym{wf}{WF}{workflow}
\newacronym{wfms}{WfMS}{Workflow Management System}
\newacronym{xes}{XES}{eXtensible Event Stream}
\newacronym{yawl}{YAWL}{Yet Another Workflow Language}
\newglossaryentry{task}{%
	name={task},description={the non-divisible, elementary activity}}

\newglossaryentry{promod}{%
	name={process model},description={the model of a process}
}
\def\LogAlph {\ensuremath{\Sigma}}
\newglossaryentry{logalph}{
	name={log alphabet},description={the process alphabet, as reflected in a log},%
	symbol={\LogAlph}}
\def\Evt {\ensuremath{e}}
\newglossaryentry{evt}{
	name={event},description={a record of an instantaneous fact during the process enactment},%
	symbol={\Evt}}
\def\Trc { \ensuremath{\tau} }

\newglossaryentry{trace}{
	name={trace},description={a sequence of \glsplural{evt}},%
	symbol={\Trc}}
\def\EvtLog {\ensuremath{L}}

\newglossaryentry{evtlog}{
	name={event log},description={a collection of \glstext{evttrace}s},%
	symbol={\EvtLog}}

\renewcommand\tabcolsep{5pt}
\newcolumntype{d}{>{\columncolor{gray!10}}c}
\newcolumntype{m}{>{\columncolor{gray!10}}l}
\newfloatcommand{capbtabbox}{table}[][\FBwidth]
\newenvironment{iiilist}%
{\begin{inparaenum}[\itshape(i)\upshape]}%
{\end{inparaenum}}

\RequirePackage{xparse}
\NewDocumentEnvironment{AuthNote}{+o+o}{%
	\IfValueT{#2}{\marginnote{\scriptsize{#2}}}%
	\begin{scriptsize}%% Change the default size of the font
		\colorbox{gray}%
		{\color{white} Note\IfValueT{#1}{ (#1)}:}%
		\quad%
		\color{brown}%% Change the colour to brown
}{%
	\normalcolor%% Restore the normal colour
	\end{scriptsize}%% Restore the normal size
}

\RequirePackage{pifont}

\RequirePackage{lipsum}
\newcommand{\LipsumGray}[1][]{{\color{gray}\ifthenelse{\equal{#1}{}}{\lipsum}{\lipsum[#1]}}}

\RequirePackage{siunitx}
\newcolumntype{D}[1]{S[
	table-omit-exponent,
	round-mode=places,
	round-integer-to-decimal,
	round-precision={#1}]} % Rounds to the given number of decimals

\providecommand{\eg}{{e.g.,}\xspace}

\providecommand{\tool}{{CONFETTY}\xspace}
\providecommand{\processInt}{{Process Interface}\xspace}
\providecommand{\confidInt}{{Confidentiality Interface}\xspace}

\newcommand{\SmallCode}[1]{\footnotesize\texttt{#1}\normalsize}

\input{addon/commands-for-inline-referenceable-objects-with-counter}
\begin{document}
\title{%
Balancing Confidentiality and Transparency for \\ Blockchain-based \\ Process-Aware Information Systems%
}
%
% If the paper title is too long for the running head, you can set
% an abbreviated paper title here
%
\titlerunning{Confidentiality and Transparency for Blockchain-based PAISs}
%
%\author{%\phantom{Hidden due to double-blindness policy}}
\author{Alessandro~Marcelletti\inst{1}\orcidlink{0000-0003-1192-6696} \and Edoardo~Marangone\inst{2}\orcidlink{0000-0002-0565-9168} \and
            Michele~Kryston\inst{3}\orcidlink{0009-0000-1491-2471} \and
		Claudio~Di~Ciccio\inst{3}\orcidlink{0000-0001-5570-0475}
}

\authorrunning{Marcelletti et al.}
% First names are abbreviated in the running head.
% If there are more than two authors, 'et al.' is used.
%
%
%\institute{\phantom{Hidden due to double-blindness policy}}
\institute{
	University of Camerino, Camerino, Italy, \email{\href{mailto:alessand.marcelletti@unicam.it}{alessand.marcelletti@unicam.it}}
	\and
 	Sapienza University of Rome, Rome, Italy, \email{\href{mailto:edoardo.marangone@uniroma1.it}{edoardo.marangone@uniroma1.it}}
 	\and
	University of Utrecht, Utrecht, The Netherlands, \email{\href{mailto:m.kryston@uu.nl}{m.kryston@uu.nl}},\email{\href{mailto:c.diciccio@uu.nl}{c.diciccio@uu.nl}}
    	%\email{\href{mailto:c.diciccio@uu.nl}{c.diciccio@uu.nl}}
}

\maketitle              % typeset the header of the contribution
%
%\vspace{-6em}
\begin{abstract}
Blockchain enables novel, trustworthy Process-Aware Information Systems (PAISs) by enforcing the security, robustness, and traceability of operations.
In particular, transparency ensures that all information exchanges are openly accessible, fostering trust within the system. Although this is a desirable property to enable notarization and auditing activities, it also represents a limitation for such cases where confidentiality is a requirement since interactions involve sensitive data.
Current solutions rely on obfuscation techniques or private infrastructures, hindering the enforcement capabilities of smart contracts and the public verifiability of transactions.
Against this background, we propose CONFETTY, an architecture for blockchain-based PAISs to preserve confidentiality and transparency. Smart contracts enact, enforce and store public interactions, while attribute-based encryption techniques are adopted to specify access grants to confidential information. 
We assess the security of our solution through a systematic threat model analysis and evaluate its practical feasibility by gauging the performance of our implemented prototype in different scenarios from the literature.

\end{abstract}
\section{Introduction}
\label{sec:introduction}
Blockchain enables new forms of trustable Process-Aware Information Systems (PAISs) and inter-organizational collaborations~\cite{stiehle_process_blockchain_review}. 
This is possible thanks to the characteristics of public permissionless blockchains, which provide strong guarantees removing the need for third-party authorities~\cite{BCopportunities}. 
Starting from the foundation, the distributed ledger and consensus mechanisms permit distributed and immutable information storage. These features, combined with the cryptographic primitives, ensure security properties and accountability of recorded operations. Smart contracts are programs immutably stored inside the blockchain that can encode and enforce business logic.
On top of this, transparency is a key aspect, as information stored in the ledger and performed operations are accessible to everyone within the system. 

Business logic enforcement and transparency are highly desirable properties for collaborative process execution and monitoring~\cite{DiCiccio.etal/SoSyM2022:BlockchainForProcessMonitoring}. 
%mnonitoring2
However, public observability of the whole set of exchanged data hinders the adoption of blockchain-based solutions in contexts where confidentiality is a critical requirement. Striking a balance between these two aspects is crucial in highly regulated sectors wherein sensitive data are treated (consider, for instance, the domains of the pharmaceutical supply-chain and healthcare~\cite{privacy_pharma_bc2}).
%privacy_pharma_bc1
Existing solutions to solve this conundrum employ private permissioned blockchains~\cite{privacy_pharma_bc3,Corradini.etal/BCRA2021:ModelDrivenEngineering}. Nevertheless, these solutions may hide data from auditors and assume high robustness and security guarantees of the consortium's system. Other approaches propose encryption techniques to restrict visibility of data stored on-chain only to authorized parties~\cite{Koepke.etal/FGCS2023:DesigningSecureBusiness,Data-Sharing-System-Smart-Cities,Marangone.etal/EDOC2023:MARTSIA,Marangone.etal/BPM2022:CAKE}.  
However, their integration with process management systems is limited or non-existent. 

Against this background, we propose an approach and an architecture for \textbf{blockchain-based process enactment that preserves the confidentiality of exchanged information while keeping public enforcement and transparency of process execution.}  
We name our approach CONFidentiality EnforcemenT TransparencY (\tool).
In particular, we rely on smart contracts to encode and execute business process logic while logging the interactions between parties, and we resort to \acrfull{maabe}~\cite{MAABE} to control the access of different parties to the activities' data payloads and information artifacts, thus safeguarding confidentiality.
We demonstrate the security of our approach against a threat model for the proposed architecture.
We implemented our artifact and evaluated the feasibility of our solution by showing its application in the context of a healthcare scenario taken from the literature. Also, we analyzed the execution performance with collaborative processes presented in the related scientific literature. 

The remainder of the paper is organized as follows. \Cref{sec:example} illustrates the example and highlights the motivation for this work. \Cref{sec:background:sota} provides the analysis of the state of the art and introduces the fundamental concepts on which CONFETTY relies. \Cref{sec:approach} describes the proposed architecture.
\Cref{sec:validation} evaluates our solution and provides an assessment of its security and feasibility. Finally, \cref{sec:conclusions} concludes the work and draws future research directions.

\section{Example, Problem Illustration and Requirements}
\label{sec:example}
\begin{figure}[tb]
	\centering
	\includegraphics[width=\linewidth]{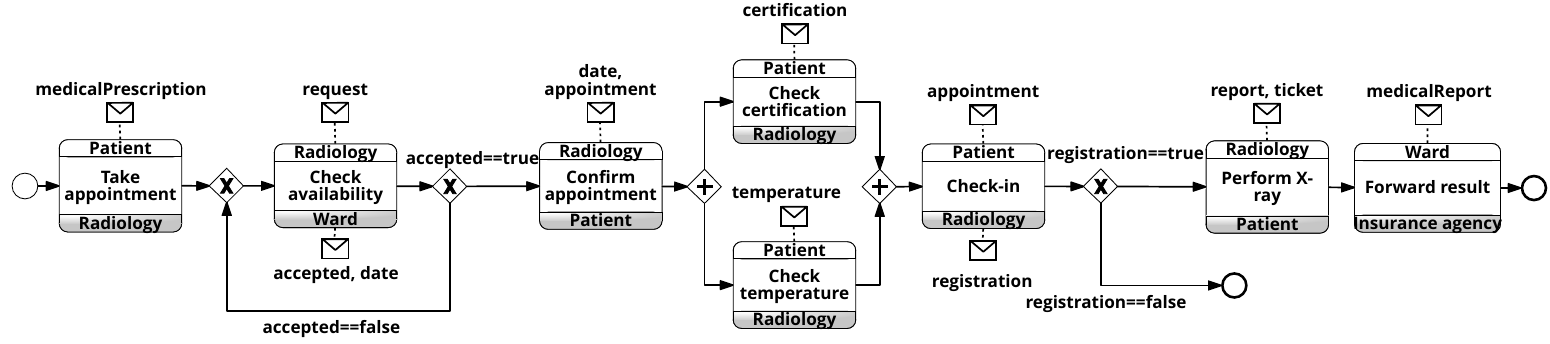}
	\caption{BPMN choreography diagram of an X-ray diagnostic analysis~\cite{runningExample}}
	\label{fig:running_example}
\end{figure}
To illustrate the problem we aim to tackle, we introduce a running example in the healthcare domain inspired by~\cite{runningExample}. \Cref{fig:running_example} depicts the management process for X-ray diagnostics in the form of a BPMN choreography diagram. 
The patient first presents the medical prescription, asking for an appointment. The radiology department checks the availability of the ward, and if a date is available, the appointment is confirmed. Otherwise, there is a new tentative. After obtaining the registration details, the temperature and vaccine certification of the patient are verified. 
A radiology clerk checks whether the appointment can be confirmed. If so, the X-ray exam is done, and the results are provided to the ward with a final report. In the end, an insurance agency receives the patient's results. The insurance agency can access the patient's medical report to follow up with compensatory actions in the subscribed contract (e.g., the reimbursement of medical bills). 
Beyond the boundaries of the choreography diagram in~\cref{fig:running_example}, we assume the presence of an inspector of the local Ministry of Health. They are not among the process participants, yet they should have full access to the information exchanged even after the conclusion of the instance for auditing purposes.

\Cref{tab:requirements} lists the requirements derived from the above use case. 
To begin with, ensuring public execution and tracking of healthcare procedures is critical to improving process efficiency and trust in the public towards the healthcare system. We concretize this aspect in~\cref{req:process}. 
Enabling auditing of those procedures is crucial since it permits investigations on misconduct or lack of compliance with regulations (\cref{req:auditability}). 
This is allowed by the data-based evidence, which should be ensured by information storage in permanent, tamper-proof and non-repudiable information artifacts (\cref{req:artifacts}). 
The guarantees offered by a public blockchain-based PAIS are fit for purpose from this standpoint.
%#R3
However, sensitive patient data must be shielded from unauthorized inspections (\cref{req:access}).
%#R4
For instance, although the insurance agency is involved in the process, it should not have full access to all exchanged data (e.g., the vaccine certification).
Publicly disclosing the result of the diagnostic analysis or information about vaccine certification would severely violate the privacy sphere of the patient. 
To prevent this, access should be restricted only to designated and authorized entities through a fine-grained access control mechanism (\cref{req:policies}).
Balancing a secure, publicly verifiable and decentralized infrastructure with the need for data confidentiality remains an open challenge, which motivates our research.

\begin{table}[tb]
	\caption{Requirements and corresponding actions in the approach}
	\label{tab:requirements}
	\resizebox{\textwidth}{!}{
		\begin{tabular}{p{0.5cm} p{5cm} p{6.8cm}} \toprule
	& \textbf{Requirement} & \textbf{Approach} \\ \midrule
    {\aReq\label{req:process}} & Public enforcement and transparency of process execution should be guaranteed & The control-flow of the process is managed by smart contracts deployed on public blockchain \\ %, ensuring enforcement of process states and since it runs on chain its execution is public verifiable \\
    {\aReq\label{req:auditability}} & The process should be independently auditable with low overhead while guaranteeing authenticity & On-chain information retains publicly available process execution tracking information and hashes with locators for associated encrypted data \\
    {\aReq\label{req:artifacts}} & Information artifacts should be written in a permanent, tamper-proof and non-repudiable way & Messages are stored in a tamper-proof distributed off-chain file storage, and the resource locators are stored on-chain to keep track of information \\
    {\aReq\label{req:access}} & Access to information artifacts should be controlled %based on the attributes that a reader bears, 
    while ensuring their overall integrity and availability & \acrshort{maabe} encrypts information artifacts stored in a distributed environment. Decryption is possible only if the requester holds the necessary attributes. \\
    {\aReq\label{req:policies}} & User-defined policies should control access levels for authenticated users  with a fine-granular scheme & \acrshort{maabe} policies are associated with messages within individual activities \\

	\bottomrule
\end{tabular}
	}
\end{table}

\section{Background and State of the Art}
\label{sec:background:sota}
In this section, we first describe the core concepts underlying our approach, namely blockchain technology and \ABE, and then illustrate how these pillars serve as the foundation for existing approaches in the literature relevant to our investigation.
\subsection{Background}
\label{sec:background}
\paragraph{Blockchain technology.}
A ledger is an append-only singly-linked list of transactions, representing asset transfers among accounts. 
A blockchain collates ledger segments into blocks, with headers storing block numbers and hashes linking to previous blocks, ensuring transaction order and data integrity.  
Selected nodes append new blocks and receive cryptocurrency as rewards. The blockchain is replicated across all nodes, ensuring transparency. Consensus protocols and cryptographic primitives enable public verification and decentralization without third-party oversight.
Several platforms implement the blockchain protocol (e.g., Ethereum~\cite{Wood/2014:Ethereum}).
Some blockchains allow for the deployment of \emph{smart contracts}~\cite{smart_contracts} which are programs deployed onto the blockchain and executed within a dedicated virtual machine (e.g., the Ethereum Virtual Machine, EVM). Smart contracts functions are invoked through transactions and their core features are code immutability and invocation storage on the blockchain, making such information available for auditing purposes~\cite{DiCiccio.etal/SoSyM2022:BlockchainForProcessMonitoring}. 
Smart contracts execution incurs a cost on the invoker. The cost is measured in gas units and depends on the complexity of the called function and the size of the exchanged data. To save on the latter, distributed tamper-proof storage systems are often employed to store the actual data in a non-modifiable file, while a permalink to that file is saved in the smart contract state. 
The InterPlanetary File System (IPFS)~\cite{IPFS} is a typical example of such a system. Based on a peer-to-peer network wherein each node stores chunks of the data, IPFS binds every file with a string used both as a unique identifier and resource locator. The string is based on the hash of the file content itself so that if the file is altered, the permalink does not match.  
As we will see next, these features fostered the development of blockchain-based PAISs using smart contracts as the backbone for business (process) rule enforcing (in our example, the BPMN choreography is translated into smart contract code, and its steps correspond to invocations to its functions)~\cite{runningExample,DiCiccio.etal/InfSpektrum2019:BlockchainSupportforCollaborativeBusinessProcesses}.

\paragraph{Attribute-based Encryption.}
\ABE is a public key encryption where the ciphertext, an encrypted plaintext, and the corresponding decryption key are linked through custom boolean attributes~\cite{ABE}. In \CPABE~\cite{CP-ABE}, a variant of ABE, every user bears a set of attributes encoded in their key. 
For example, the owner of the blockchain account \verb|0xB0|\ldots\verb|1AA1| can be endowed with attributes \texttt{Patient} and \texttt{PID476948} specifying role and process instance \num{476948}, respectively.
Data owners write policies to be attached to the \CPABE-encrypted document (the \emph{ciphertext}). Those policies express conditions for granting in-clear access to the data. Policies are propositional logic formulae employing {\ABE} attributes as literals. The policy is evaluated on attributes associated with a potential reader: Only if the formula is satisfied can the user's key decrypt the data.
For instance, the key of the user \verb|0xB0|\ldots\verb|1AA1| can decrypt data locked with a policy like \texttt{PID476948 and PATIENT}, but cannot unlock the contents of a document associated with a policy like \texttt{PID476949 and RADIOLOGY}.
Attributes are typically assigned by one authority: \MAABE~\cite{MAABE} is a variant of \ABE that removes this single point of failure using a multi-authority method. In \MAABE, every authority creates a part of the decryption key.
Once collected the user merges all parts to obtain the final key. In this paper, we employ the Ciphertext-Policy variant of \MAABE~\cite{MultiAuthorityCP}. It integrates \MAABE with the aforementioned policy-based approach.

\subsection{State of the Art}
\label{sec:sota}
Over the last few years, several solutions that automate collaborative processes using blockchain technology have been designed~\cite{stiehle_process_blockchain_review,DiCiccio.etal/InfSpektrum2019:BlockchainSupportforCollaborativeBusinessProcesses,Lopez-Pintado.etal/SPE2019:Caterpillar,Tran.etal/BPMDemos2018:Lorikeet}. 
%Several works in this field have
Such approaches demonstrated the effectiveness of blockchain-based solutions in improving trust among participants in multi-party collaborations and allowing for monitoring and auditing~\cite{DiCiccio.etal/SoSyM2022:BlockchainForProcessMonitoring,Corradini.etal/ACMTMIS2022:EngineeringChoreographyBlockchain}.
%enhanced with model-driven approaches, Several works in this field have demonstrated the effectiveness of blockchain-based solutions in improving trust among participants in multi-party collaborations~\cite{Weber.etal/BPM2016:UntrustedBusinessProcessMonitoringandExecutionUsingBlockchain} even in adversarial settings~\cite{Madsen.etal/FAB2018:CollaborationamongAdversaries:DistributedWorkflowExecutiononaBlockchain}, enhance verifiability of workflows with model-driven approaches~\cite{Lopez-Pintado.etal/SPE2019:Caterpillar,Tran.etal/BPMDemos2018:Lorikeet}, allow for monitoring~\cite{DiCiccio.etal/SoSyM2022:BlockchainForProcessMonitoring}, mining~\cite{Klinkmueller.etal/BCForum2019:ExtractingProcessMiningDatafromBlockchainApplications}, security~\cite{Koepke.etal/FGCS2023:DesigningSecureBusiness}, and auditing~\cite{Corradini.etal/ACMTMIS2022:EngineeringChoreographyBlockchain,Corradini.etal/BCRA2021:ModelDrivenEngineering}.
These studies enhance the integration of blockchain technology with process management, unlocking security and traceability benefits. However, they primarily focus on the control-flow perspective and lack mechanisms for secure access control to the stored data on public platforms.

A relevant area of research for our work pertains to data privacy and integrity within blockchain systems. Several studies examine the application of encryption techniques to achieve this goal.
Hawk~\cite{Hawk} employs user-defined private smart contracts to automate the implementation of cryptographic protocols.
%Abid et al.~\cite{Abid2024privacytraceabilityssibpm} leverage Zero-Knowledge Proof (ZKP) and Fully Homomorphic Encryption (FHE) to build a solution that supports privacy-preserving traceability and the analysis of encrypted data. They store on-chain only the hash of the data used to verify the correctness, while all the computations and the data are stored off-chain.
%Benhamouda et al.~\cite{BenhamoudaCanAP} use a public blockchain storage layer for confidential data.
%Peng et al.~\cite{peng2023peer} propose a consortium blockchain as a decentralized storage system for file sharing among organizations. 
%Wu et al.~\cite{nannawu} present an attribute-based access control scheme that preserves the privacy of both attributes and policies.
Pham et al.~\cite{B-Box} propose a decentralized storage solution that integrates \IPFS, \MAABE, and blockchain technology.
Hong et al.~\cite{Linjian} introduce a data-sharing solution based on decentralized ABE (\MAABE), blockchain, and IPFS.
Bramm et al.~\cite{secrypt18} present BDABE, an access control mechanism that applies a distributed ABE scheme within a consensus-driven infrastructure for real-world applications.
%Feng et al.~\cite{info14050281} combine attribute-based encryption ABE and identity-based encryption (IBE) to achieve efficient data sharing and verification of data correctness. They employ blockchain technology to ensure tamper-proof and regulated data storage.
Yan et al.~\cite{Yan2023} introduce a scheme for fine-grained access control by implementing proxy encryption and decryption while supporting policy hiding and attribute revocation. The encrypted data is stored on \IPFS, and the metadata is stored on the blockchain.
All these studies, along with the aforementioned~\cite{Marangone.etal/BPM2022:CAKE,Marangone.etal/EDOC2023:MARTSIA,marangone2023enabling}, cater for secure access control on blockchains but lack the integration of this technology with process management systems. 

To the best of our knowledge, this is the first work that focuses on combining blockchain-based process execution engines with encryption schemes that guarantee data confidentiality and access control over data shared on public blockchain platforms. In the next section, we provide a description of our solution.

\section{The \tool Architecture}
\label{sec:approach}
In this section, we describe the proposed \tool architecture. 
We first introduce its core components and functionalities (\cref{sec:arch:functional}). Then, we describe how components interact and manage information to support confidential-preserving blockchain-based process execution while meeting \cref{req:access,req:artifacts,req:auditability,req:policies,req:process} (\cref{sec:architecture:workflow}). Due to space restrictions, we do not enter the technical details of our approach but provide an overarching presentation of our solution's rationale and key concepts, motivating them with the requirements they aim to meet.
% We conclude the section with some considerations on the benefits brought by our approach of merging blockchain technologies and \MAABE.

\subsection{Functional Viewpoint}\label{sec:arch:functional}
\begin{figure}[tb]
	\centering
\includegraphics[width=\textwidth]{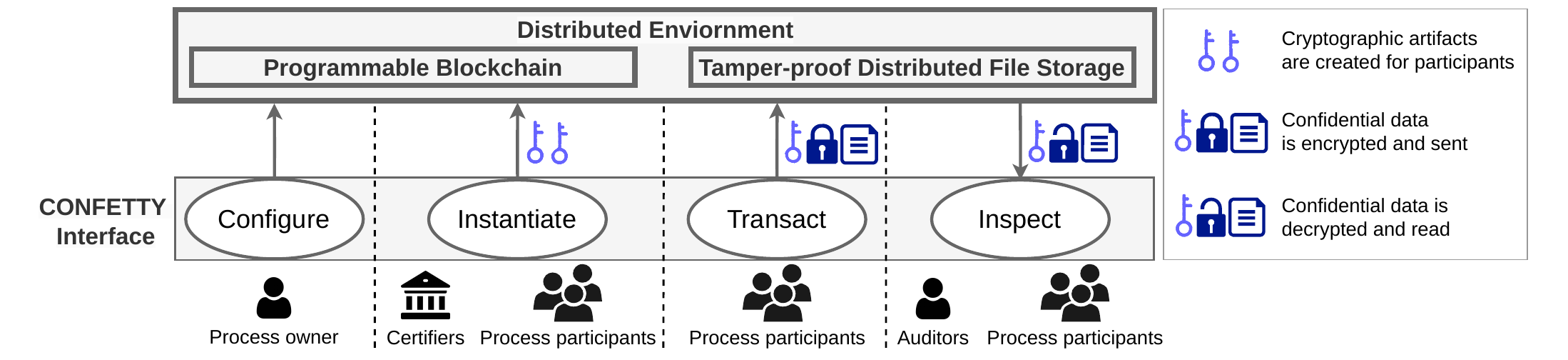}
	\caption{A graphical sketch of the \tool functionalities, users, artifacts, and interfaces}
	\label{fig:functionalities}
\end{figure}
\Cref{fig:functionalities}
provides an overview of the \tool approach,
which caters for the following main functionalities. 
\begin{iiilist}
\item The {\textbf{configure}} functionality helps the process owners specify the process to run on-chain. It requires the registration of process information, including activities to execute, data to exchange, control-flow routing, the expected participants' roles (like the information enclosed in the choreography diagram in \cref{fig:running_example}) and how they can operate with confidential data (via {\MAABE} policies) to control information access. 
\item The {\textbf{instantiate}} functionality lets certifiers authenticate the participants, define their role in the process (e.g., that user \verb|0xB0|\ldots\verb|1AA1| is a patient), and kickstart a new process instance on-chain by setting its public state. 
\item The {\textbf{transact}} functionality allows process participants to interact, following the process, to exchange data at run time. This functionality manages both public and confidential data, updating and enforcing the process state according to the logic in the blockchain. 
\item The {\textbf{inspect}} functionality allows participants and auditors to access public and confidential data. 
\end{iiilist}

\begin{figure}[tb]
        \centering
    	\includegraphics[width=0.75\linewidth]{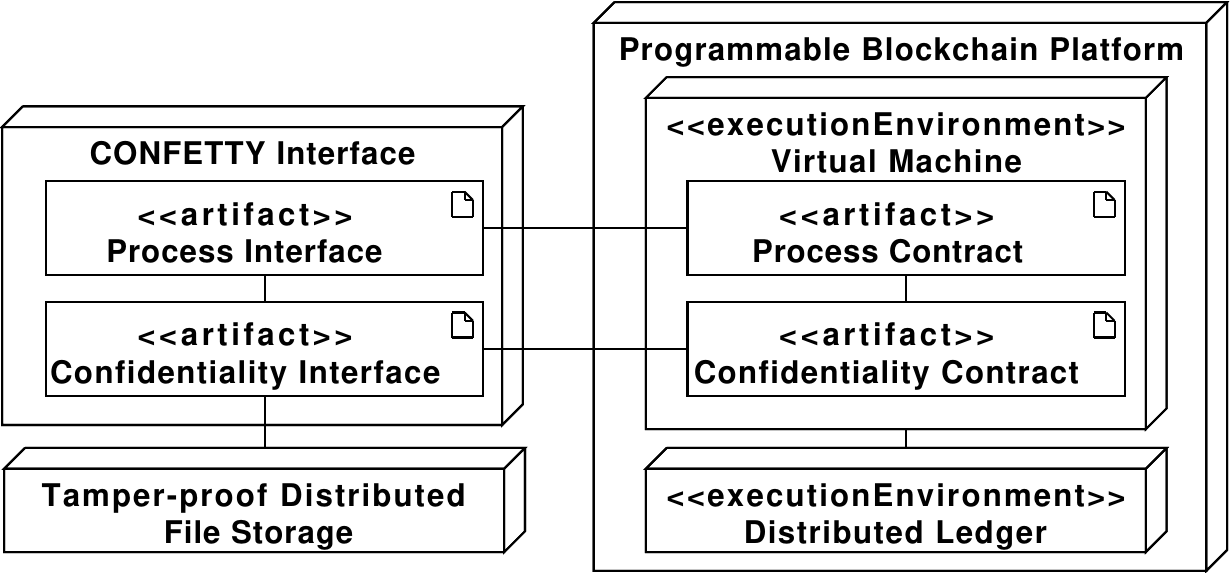}
    	\caption{The deployment diagram of \tool}
    	\label{fig:architecture}
\end{figure}
%\end{wrapfigure}
%
The functionalities above are realized by the \tool main components depicted in \cref{fig:architecture}. %:
\tool relies on two external components used as architectural buttresses:
\begin{iiilist}
\item A \textbf{{Programmable Blockchain Platform}} (like Ethereum),
\item and a \textbf{{Tamper-proof Distributed File Storage}} (like IPFS). 
\end{iiilist}
%In addition to the above inner components, the architecture relies on two external ones: a \textbf{{Programmable Blockchain Platform}} (like Ethereum) and a \textbf{{Tamper-proof Distributed File Storage}} (like IPFS). 
%BLOCKCHAIN
The former maintains the public state, allowing for the creation of logic and constraints while notarizing the executed operations. It comprises two core elements: The {Distributed Ledger}, storing the transactions, and the {Virtual Machine}, hosting and running smart contracts.
%IPFS
The latter (which we will henceforth name as Distributed FS for brevity) stores business data and thus improves the performance and costs of a blockchain as in common practice. The data is saved on the external storage (upon encryption, if it must remain secret), while a resource locator with the hash of that data is kept on-chain for notarization and future retrieval. Notice that we take inspiration from known patterns adopting Distributed FSs for large data storage~\cite{Xu2019}. Based on this known pattern, we devise new techniques to preserve secrecy and control access to that process data at runtime.
%combining xxx mechanisms and public process management. 
%
%To interact with the mentioned components, our architecture relies on the \tool Interface, containing the purpose-specific artifacts providing utility and communication capabilities with the aforementioned components. Specifically, the 
The primary components of \tool are the following:
\begin{iiilist}
\item The \textbf{{\processInt}} handles all the operations involving the participants and related to the process. This is done by acting as an interface for the blockchain, abstracting from the technical implementation and exposing functionalities to the users.
In particular, the \processInt instantiates the process on the blockchain on given process specification data. To support the \emph{instantiate} functionality, it provides capabilities for role assignment to participants. % and their identities. 
\item The \textbf{{Process Contract}} is a smart contract that manages the process instances and their execution. It elaborates and manages all process specifications and their instances' public states, acting as a software realization both of factory and proxy patterns~\cite{Xu2019}. 
%~\cite{gamma1995design} akin to~\cite{Corradini.etal/ACMTMIS2022:EngineeringChoreographyBlockchain}. 
It supports the \emph{instantiate} functionality by enforcing the control flow, the data flow, the assignment of tasks to participants, and the routing of decision points.
In this way, the Process Contract updates the public state of the process and generates transparent records that are accessible and verifiable by interested auditors.
\item The \textbf{{\confidInt}} is in charge of handling the participants' authorizations and the operations over confidential data. 
It operates as a fa\c{c}ade towards the blockchain to store access grants, distributes decryption keys to the users solely based on their roles and involvement in the process, and handles the storage and retrieval of encrypted data.
\item The \textbf{{Confidentiality Contract}} is a smart contract handling the notarization of authorizations for confidential data, as well as for their writing and access. 
\end{iiilist}

We conclude this subsection with a few considerations about the users and trust model. We assume that the users we mentioned while discussing the functionalities (i.e., process owners, certifiers, process participants, and auditors) possess a blockchain account with a key pair to be able to publicly identify themselves (as with the aforementioned user \verb|0xB0|\ldots\verb|1AA1|) and sign blockchain transactions built through the {\tool} modules. We also remark that we consider the {\tool} platform as a trusted party, while users are assumed to be \emph{honest but curious}. Finally, notice that the involvement of a pre-appointed third-party certifier to authenticate process participants is necessary as a user alone can confirm their identity (e.g., via the aforementioned public-private key scheme) but cannot attest alone to the truthfulness of assertions regarding themselves. The selection and appointment of certifiers transcends the goals of this paper, but in our motivating scenario, e.g., certifiers can be trusted third parties like the Public Ministry of Health (for hospital personnel), the Public Registry (for citizens), and the Insurance Registrar (for agencies), or subsidiaries thereof. In our architecture, certifiers operate like blockchain oracles~\cite{DBLP:conf/bpm/MuhlbergerBFCWW20}.

\subsection{Information Viewpoint}\label{sec:architecture:workflow}
After providing an overview of the \tool architecture, we describe how information is handled in realizing the aforementioned functionalities by the \tool components. 

\begin{sloppypar}
%\Cref{fig:design} depicts the 
\begin{table}[tb]
	\centering
	\resizebox{0.66\textwidth}{!}{%
		\begin{tabular}{|c|c|}
\toprule
\textbf{Message} & \textbf{Access policy} \\ \midrule
Medical prescription & \SmallCode{MINISTRY-INSPECTOR or (\$PID and (PATIENT or RADIOLOGY))} \\ 
Registration & \SmallCode{MINISTRY-INSPECTOR or (\$PID and (RADIOLOGY or PATIENT))} \\ 
Report & \SmallCode{MINISTRY-INSPECTOR or (\$PID and (RADIOLOGY or WARD))} \\ 
%Medical report & \texttt{MINISTRY-INSPECTOR or (\$PID and (INSURER or WARD))} \\ 
\bottomrule
\end{tabular}%

	}
	\caption[Access policies]{Access policies. \SmallCode{\$PID} represents a placeholder for the identifier of the process instance before its assignment}
	\label{tab:access-policies}
\end{table}
The first functionality is \emph{configure}. Here the process owner sends the process specification files (like the diagram in \cref{fig:running_example}) to the \processInt. %(1.0).%(1.1) 
The \processInt, in turn, processes the input data, storing the behavioral and business constraints it reports. 
%Also, (1.2) it builds a policy template, associating roles of the process with the data they can access via encryption directives. Once created, the policy template is returned to the process owner. 
%In our running example, an extract of a policy template is:\\
%(1.2)
Also, it builds a set of parametric access policies, associating roles of the process with the data they can access via encryption directives (meeting \cref{req:policies}). Once created, the policies are returned to the process owner. 
The one associated with the medical prescription (sent by the patient to the radiology clerk to take an appointment in \cref{fig:running_example}), e.g., reads
``\SmallCode{\$PID and (PATIENT or RADIOLOGY)}''.
%In our running example, an extract of a policy is:\\
%\SmallCode{\{medicalPrescription:(\$PID and (PATIENT or RADIOLOGY)),
%registration:\\ (\$PID and (RADIOLOGY or PATIENT)), report:(\$PID and (RADIOLOGY or WARD))\}}.
%Here \SmallCode{medicalPrescription}, \SmallCode{registration}, and \SmallCode{report} represent messages, while \SmallCode{PATIENT}, \SmallCode{RADIOLOGY} and \SmallCode{WARD} are the roles of users who can access them. %Certifying entities that have to attest those roles are specified as @AUTH1, @AUTH4, @AUTH2 respectively.
Here, \SmallCode{PATIENT} and \SmallCode{RADIOLOGY} are the attributes representing the actors' roles in the process.
Not all patients and radiology clerks should access any medical prescription.
Hence \SmallCode{\$PID}, an attribute placeholder for the process instance identifier (a piece of information that can be known only at runtime),
restricting access to the sole actors involved in that process instance.
We give the process owner the option to revise the policies by adding custom roles that are not expected active parties in the process but need access to data (as per \cref{req:auditability}). This is the case, for instance, of external auditors who need to certify data and its compliance with norms, laws and regulations.  
% In this case, each message is attached to a policy with an ID representing the process instance and the authorized role. 
Considering the medical prescription, the revised policy specifies that a user has to be attested to have the role of a \SmallCode{MINISTRY-INSPECTOR}, or again to be a \SmallCode{PATIENT} or \SmallCode{RADIOLOGY} within the running process instance:
``\SmallCode{MINISTRY-INSPECTOR or (\$PID and (PATIENT or RADIOLOGY))}''.
\Cref{tab:access-policies} shows an extract from the final access policies of our running example.
%Policies: {"prescription": "(476948@AUTH1 and 476948@AUTH2 and 476948@AUTH3 and 476948@AUTH4) and (RADIOLOGY@AUTH1)", "appointmentId": "(476948@AUTH1 and 476948@AUTH2 and 476948@AUTH3 and 476948@AUTH4) and (PATIENT@AUTH2)", "requestId": "(476948@AUTH1 and 476948@AUTH2 and 476948@AUTH3 and 476948@AUTH4) and (WARD@AUTH4)"}
% (1.3) (1.4) (1.5)
%
Once policies are confirmed, they are sent to the \confidInt, which stores them in the Distributed FS. Their permalink is passed back to the \confidInt to save it. 
\end{sloppypar}

\begin{figure}[tb]
    \centering
    \includegraphics[width=\linewidth]{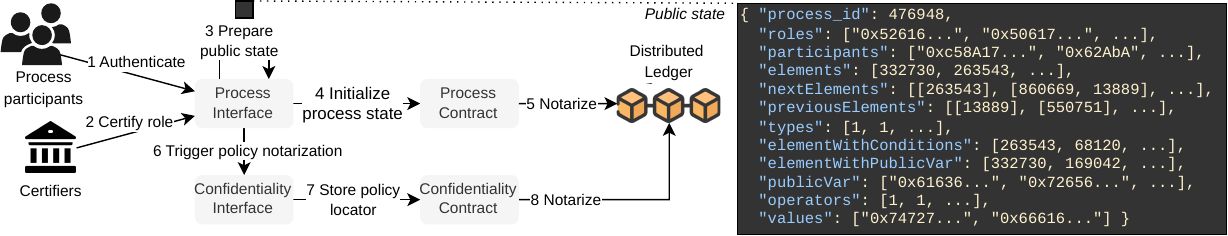}
    \caption[Communication diagram of the \emph{instantiate} functionality]{Communication diagram of the \emph{instantiate} functionality, with a snapshot of public state}
    \label{fig:enrollment}
\end{figure} 

%FASE 2
\Cref{fig:enrollment} depicts the \emph{instantiate} functionality. 
Differently from the \emph{configure}, needed once for each process, this one is designed for every new instantiation thereof. To initialize a new process instance (e.g., number 476948), participants send their data to the \processInt (1) to authenticate themselves as participants in the process with their role (e.g., that user \verb|0xc5|\ldots\verb|Fd43| operates as a \SmallCode{WARD} clerk in a new process instance \SmallCode{PID476948}). (2) The authentication is attested to by certifiers confirming the user's self-declared roles. 
When all process participants are identified and associated with a role, (3) the \processInt merges the received information with the process data acquired in the configuration. Thereupon, (4) a selected participant kickstarts a new process instance through the \processInt associated with the aforementioned information on the Process Contract, thereby initializing the instance's public state. This step is done by sending a transaction to the Process Contract with the public state of the process, including conditions driving the control flow, such as exclusive gateway choices (as per \cref{req:process}).  Notice that this crucial step is notarized on chain (5) (meeting \cref{req:auditability}). The selection of the participant, or participants, on behalf of whom the transaction is signed, transcends the scope of this paper and depends on the decision process and guidelines of the consortium. 
After initializing the process state, (6) the \processInt triggers the \confidInt, which (7) stores the policy locator previously saved on the Confidentiality Contract. Finally, (8) the \confidInt notarizes the operation.

\begin{figure}[tb]
    \centering
    %\subfloat[Public case]
    \includegraphics[width=\textwidth]{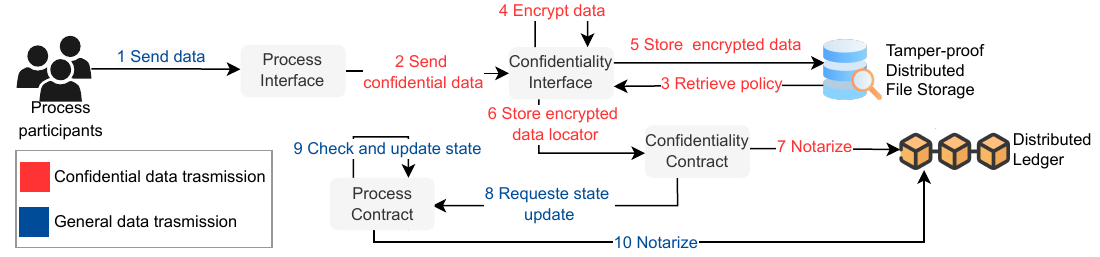}    
    \caption{Communication diagram of the \emph{transact} functionality}
    \label{fig:write}
    % \label{fig:write}  
\end{figure}

%FASE3
\Cref{fig:write} displays the \emph{transact} functionality, which represents the exchange of information during the execution of an activity in the process instance. It encompasses a public and a confidential case, depending on the process participants and their willingness to classify information as disclosable or not. 
In both cases, this functionality updates the public state in the blockchain according to previous enforcement as per \cref{req:process} (e.g., activity to execute, expected participant) and manages accessibility (meeting \cref{req:access}). 
% in case of public and confidential cases 

Let us begin with the exchange of public data (and the update of the public state). The main steps are highlighted in blue in \cref{fig:write} and salient on-chain operations are reported in \cref{alg:processContract}. 
Initially, a user sends the execution data to the \processInt in the context of an activity (\cref{req:process}). 
For example, the \SmallCode{WARD} clerk identified by \verb|0xc5|\ldots\verb|Fd43| replies to the
``Check availability'' activity request by the \SmallCode{RADIOLOGY} user within process instance {476948}. 
To do so, the clerk sends a ``Check availability'' response message wherein the \SmallCode{accepted} variable is assigned with \SmallCode{true} since the appointment is confirmed alongside an available \SmallCode{date}. The boolean answer (\SmallCode{accepted}) is public data that is used later to make control-flow decisions and advance the state of the process, defining the next activity to perform. In our scenario, the exclusive choice gateway is based on the value of the \SmallCode{accepted} variable. %3.1a(2) c
Then, the \processInt invokes the Process Contract to request a state update, sending the execution data (Line~\ref{cod:invokeUpdate}). 
%3.2a
In turn, the Process Contract enforces the business and control-flow logic through the run of its routine (based on previously agreed upon and immutable code, see \textit{configure} functionality), which checks (Line~\ref{cod:senderAndEnabled}) if the executed activity is enabled and if the message sender (in this case, user \verb|0xc5|\ldots\verb|Fd43|) is the expected one (a \SmallCode{WARD}), finally updating the state (Line~\ref{cod:updateState}) if the verification outcome is positive.
This step advances also the state of the process (Line~\ref{cod:activateNext}), updating the expected next element (here, ``Confirm appointment'') and participant (\SmallCode{RADIOLOGY}). In case the element is a gateway, its logic is automatically executed to advance the control flow of the process (Line~\ref{cod:executeGateway}). In the end, this piece of information is notarized into the Distributed Ledger (meeting \cref{req:auditability}).

\begin{algorithm2e}[tb]
	\caption{On-chain logic for public state update and next element activation}
	\label{alg:processContract}
\scriptsize
\DontPrintSemicolon
\SetKwFunction{setState}{updatePublicState}
\SetKwFunction{setStatus}{activateNext}
\SetKwProg{Fn}{Function}{:}{}

\Fn{\setState{instanceID, messageID, variables}}{ \label{cod:invokeUpdate}
authUser $\gets$ instances[instanceID].elements[messageID].role \;
element $\gets$  instances[instanceID].elements[messageID] \;
  \If{sender is authUser \& element.type is MESSAGE \& element.state is ENABLED}{ \label{cod:senderAndEnabled}
    \lForEach{variable in variables}{
      instances[instanceID].publicVars[variable.name] $\gets$ variable.value \label{cod:updateState} 
    }
    \setStatus\!(instanceID,  messageID)\;  \label{cod:activateNext}}}

\Fn{\setStatus{instanceID, messageID}}{
  currentElement $\gets$ instances[instanceID].elements[messageID]\;
  instances[instanceID].elements[messageID].state $\gets$ COMPLETED\;
  \ForEach{nextElement in currentElement.nextElements}{
    \lIf{nextElement is GATEWAY}{
      executeGateway(instanceID, nextElement.ID) \label{cod:executeGateway}
    }
    \lElse{
      nextElement.state $\gets$ ENABLED
    }
  }
}
\end{algorithm2e}
For the confidential case, we refer again to \cref{fig:write} (see the marks in red). Overall, (1) once the {\processInt} has received the data, (2) it forwards it to the \confidInt. The latter (3) retrieves the policy from the Distributed FS, and (4) uses it to encrypt the data via \MAABE (as per \cref{req:access}). For example, considering \cref{fig:running_example} and \cref{tab:access-policies}, the medical prescription of the patient is encrypted with the following policy: \SmallCode{\{MINISTRY-INSPECTOR or (PID476948 and (PATIENT or RADIOLOGY))\}}. 
In particular, the \confidInt retrieves the specific authorization for the data to be stored and embeds it in the encryption mechanism, ensuring that only authorized participants can access it. This step is needed only in confidential cases, as in public ones, information is directly stored on the blockchain and intentionally kept publicly readable. 
Once encrypted, (5) the \confidInt stores data in the Distributed FS, (6) its locator in the Confidentiality Contract, and (7) notarizes the operation in the Distributed Ledger (meeting \cref{req:auditability} and \cref{req:artifacts}).
In this case, the resource locator that contains the encrypted \SmallCode{medicalPrescription} (\eg \verb|LXBw|\ldots\verb|2DAx|) is \verb|Qmo9|\ldots\verb|0weI|.
The execution of this routine by the Confidentiality Contract includes a call to the Process Contract (8) to request a state update enclosing the execution data. Following the example, this data is the activity being executed (``Check availability''), the user that performs it (\eg, \SmallCode{PATIENT}, along with their corresponding address \verb|0x2e|\ldots\verb|6dd9|), and the process instance (\eg {476948}).
Then, (9) the Process Contract checks the current public state to update it. The operations involving the smart contracts (10) are notarized in the Distributed Ledger.

The \emph{inspect} functionality pertains to the access to message data. The public data can be freely retrieved, while the confidential data is restricted to the sole authorized users included in the access policy provided during the encryption phase (see the \emph{configure} functionality above) (\cref{req:policies}). 
To read public data, a process participant or an auditor may send a request to the \processInt, which in turn retrieves the necessary information from the Process Contract  
and sends it back to the requester. 
This is the case, e.g., for the \SmallCode{accepted} field of the ``Check availability'' task.
Reading confidential data requires more passages and a preliminary step. %illustrated in \cref{fig:read},
All users should request at least once (and at every attribute update) a decryption key, to be used for every message they are going to read. % (\cref{fig:request_key}) (\cref{fig:read_conf})
For the key request, the process participant (e.g., a \SmallCode{WARD} clerk) or an auditor (e.g., a \SmallCode{HEALTHCARE-INSPECTOR}) ask for a decryption key to the \confidInt. %(4.0a)
The latter forwards the request to the Confidentiality Contract, which is thus notarized on the Distributed Ledger. %(4.1a)(4.2a) 
Then, the \confidInt generates the attribute-based key (or \textit{a-b key} for short) based on the attributes of the user requesting it (e.g., \SmallCode{WARD} and \SmallCode{PID476948}, or \SmallCode{HEALTHCARE-INSPECTOR}) via \MAABE, and sends it back. %(4.3a) (4.4a) Another crucial step is the 
The aforementioned step enables the decryption of confidential data. % is depicted in \cref{fig:read_conf}.
Once the process participants have received the a-b key, they can use it to decrypt the confidential data. To do so, a user makes a request to the \confidInt. % (4.0b)
Then, the \confidInt retrieves the encrypted data from the Distributed FS and forwards it to the user. %(4.1b)(4.2b) 
At this point, the latter uses the a-b key to decrypt the confidential data via \MAABE. Only if the user's attributes satisfy the policy previously used to encrypt the data can the user's a-b key decrypt the confidential data (and the user inspects it in clear). %(4.3b)
\paragraph{A note on decryption keys.} We remark that an a-b key stays with the user and lasts for as long as their attributes are not updated (if a user is involved in a new process instance, say 476949, a new attribute gets associated with them: \SmallCode{PID476949}). The same key will be used to access all messages. For example, consider user \verb|0xB0|\ldots\verb|1AA1| (who holds the attributes \SmallCode{PID476948} and \SmallCode{PATIENT}).
Per the policies in \cref{tab:access-policies}, their a-b key can decrypt the medical prescription and the registration, but not the report.
The a-b key of the user \verb|0xc5|\ldots\verb|Fd43| (\SmallCode{PID476948} and \SmallCode{WARD}) can decrypt the report instead, but not the other two documents.
Notice that a \SmallCode{HEALTH-INSPECTOR}, instead, can use their unique a-b key to access all the aforementioned documents regardless of the process instance they pertain to.
The key, in other words, is associated with the user, and it is the sole instrument to access a document.
Therefore, the \confidInt does not need to generate a different decryption key for every document and give the same copy to multiple users, nor create numerous copies of the same document for different decryption keys (one per user), thus saving unnecessary data replications and key distributions, and ensuring integrity (\cref{req:access}). Also, as the key stays with the users, and the documents are stored and notarized on tamper-proof external infrastructures (the Distributed FS and Blockchain Platform, respectively), the information remains available even should the \tool system stop functioning (\cref{req:artifacts}). These aspects have an impact on security considerations, as we discuss next.

\section{Evaluation}
\label{sec:validation}
To assess the feasibility of the \tool architecture, we perform a two-step evaluation. First, we analyze the \tool threat model, evaluating it against the proposed architecture. Then, we describe the implemented prototype and analyze its performance. The source code of \tool, alongside the experimental data we use and performance analysis results, are available at \href{https://doi.org/10.5281/zenodo.15482587}{\nolinkurl{https://doi.org/10.5281/zenodo.15482587}}. 

\subsection{Threat Model}
\label{sec:threat-model}
\begin{wrapfigure}[14]{r}{0.5\textwidth}
        \centering
    	\includegraphics[width=\linewidth]{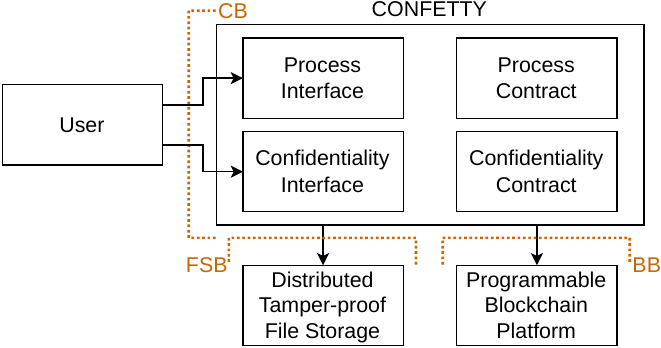}
    	\caption[The CONFETTY threat model]{The CONFETTY threat model, based on the components in \cref{fig:architecture}}
    	\label{fig:threat-model}
\end{wrapfigure}
In this section, we identify potential security threats that could compromise data availability (\cref{req:auditability}) and confidentiality (\cref{req:policies}) within our solution, and outline how the \tool architecture addresses them. %These threats establish the security requirements that our solution must satisfy.
Our threat analysis is based on the STRIDE framework~\cite{ScandariatoWJ15}, a theoretical model that categorizes threats into six distinct groups: 
spoofing (i.e., impersonation of a legitimate entity), 
tampering (i.e., unauthorized modification of data to compromise its integrity), 
repudiation (i.e., the denial of having performed a particular action), 
information disclosure (i.e., the unauthorized exposure of sensitive data), 
denial of service (DoS, i.e., the disruption or degradation of system availability), and 
elevation of privileges (i.e., the unauthorized acquisition of higher-level access rights).
We assume users and components to be honest but curious and trusted, respectively. The possible threats thus reside in the communication between them, subject to potential attacks (see~\cref{sec:arch:functional}). \Cref{fig:threat-model} displays a Data Flow Diagram (DFD), in which the orange dotted lines indicate the trust boundaries crossing data exchange points between different perimeters, represented by black arrows connecting different elements. The opening side of each boundary indicates the trusted component in the information exchange. 

The \textbf{\tool Boundary (CB)} indicates that \tool is the trusted component in the communication with a user (let it be a certifier, a process participant or owner, or an auditor). In this case, a spoofing attack could result in a user impersonating another. Still, this attack is prevented since we assume that each user is linked to the Blockchain Platform credentials. As for tampering, trust is based on the fairness of the user who sends the data. It cannot be the case for a user to repudiate an action because all of them are recorded on the blockchain with a transaction signed with the user's private key. An attacker could sniff the data a user sends to \tool. 
However, we resort to TLS channels for those communications, specifically designed to prevent information leakage~\cite{Thomas2000SSLAT}. DoS attacks are, in principle, feasible unless we pre-filter the range of potential clients invoking the \tool platform. Indeed, all functionalities foresee an initial data input from users (see, e.g., step 1 in \cref{fig:write})
which malicious nodes could exploit with multiple calls and bulky payloads. However, DoS attacks would lead to a disruption or degradation of the system's availability. Yet, all data, state, and other process instances' information can be retrieved from the Blockchain Platform and the Distributed Tamper-proof File Storage to reactivate or replicate the system without any data loss. Information disclosure is prevented because all sensitive information is encrypted via \MAABE. Elevating a user's privileges is impossible since attributes and permissions are linked to a blockchain account.

The Blockchain Platform is the trusted component in the \textbf{Blockchain Boundary (BB)}. In this case, a spoofing attack could result in a malicious version of \tool impersonating the original software object. This attack is prevented since \tool operates as a gateway for transactions ultimately signed by users via their private keys. 
Information disclosure can be achieved only on public data or resource locators of sensitive data. However, the latter is stored upon \MAABE-encryption, making its content unreadable by malicious users. Other possible attacks (DoS, tampering, repudiation, elevation of privileges) are counteracted by the design of blockchain protocols.
As for the Distributed Tamper-proof \textbf{File Storage Boundary (FSB)}, tampering or DoS attack and elevation of privileges are counteracted by design, too. Spoofing and repudiation attacks are ineffective since the authorship of the provided information is proven by signatures of transactions stored on-chain. Since we resort to \MAABE encryption, information disclosure is negated since the sensitive data stored is encrypted.

\subsection{Implementation}
\label{sec:implementation}
\tool is deployed on three main tiers. An application server hosts the \confidInt and the \processInt, developed in Python 3.6.9 and Java SDK-13, respectively.
Users communicate with it via TLS communication channels~\cite{Thomas2000SSLAT}. 
We utilize IPFS\footnote{IPFS: \href{https://ipfs.tech/}{\nolinkurl{ipfs.tech/}}; Sepolia: \href{https://sepolia.dev/}{\nolinkurl{sepolia.dev}}; Ganache: \href{https://archive.trufflesuite.com/ganache/}{\nolinkurl{archive.trufflesuite.com/ganache}}; Web3.js: \href{https://web3js.readthedocs.io/en/v1.10.0/}{\nolinkurl{web3js.readthedocs.io/en/v1.10.0/}}. Accessed: \today. \label{foot:technologies}} as the Distributed Tamper-proof File Storage. We employ Sepolia and Ganache as Programmable Blockchain Platforms with their Ethereum Virtual Machine (EVM) as a test environment and local RPC, respectively, connected via Web3.js\textsuperscript{\ref{foot:technologies}}. 

\subsection{Execution Analysis}
\label{sec:performance}
%To provide a preliminary analysis of the practical feasibility of \tool, we evaluate the implemented prototype at work, discussing its performance in terms of time and costs. 
%To analyze the practical feasibility of \tool, 
Motivated by \cref{req:auditability}, we evaluated our prototype %, discussing its performance 
in terms of time and cost according to the four key functionalities of \tool (see \cref{sec:arch:functional}). 
%For this purpose,  and scalability 
%We gauged \tool while operating with our use case scenario (\cref{sec:example}) and two other processes from the literature about retail~\cite{Corradini.etal/BCRA2021:ModelDrivenEngineering} and incident management~\cite{Corradini.etal/ACMTMIS2022:EngineeringChoreographyBlockchain}. 
We set the number of {\MAABE} authorities to four and performed the experiments on a machine equipped with a processor Intel(R) Core(TM) i7-13700H CPU @ \qty{2.40}{\giga\hertz} with \qty{14} cores and \qty{32}{\giga\byte} of RAM.
To isolate the architecture functionalities, we excluded human interactions and the time needed for blockchain-specific operations. %like finalizing blocks on the distributed ledger. 
Such operations strictly depend on the blockchain protocol and the platform adopted, representing technical choices that are out of the scope of this work.

\begin{figure}[tb]
	\subfloat[Time analysis]{{\includegraphics[width=0.5\textwidth]{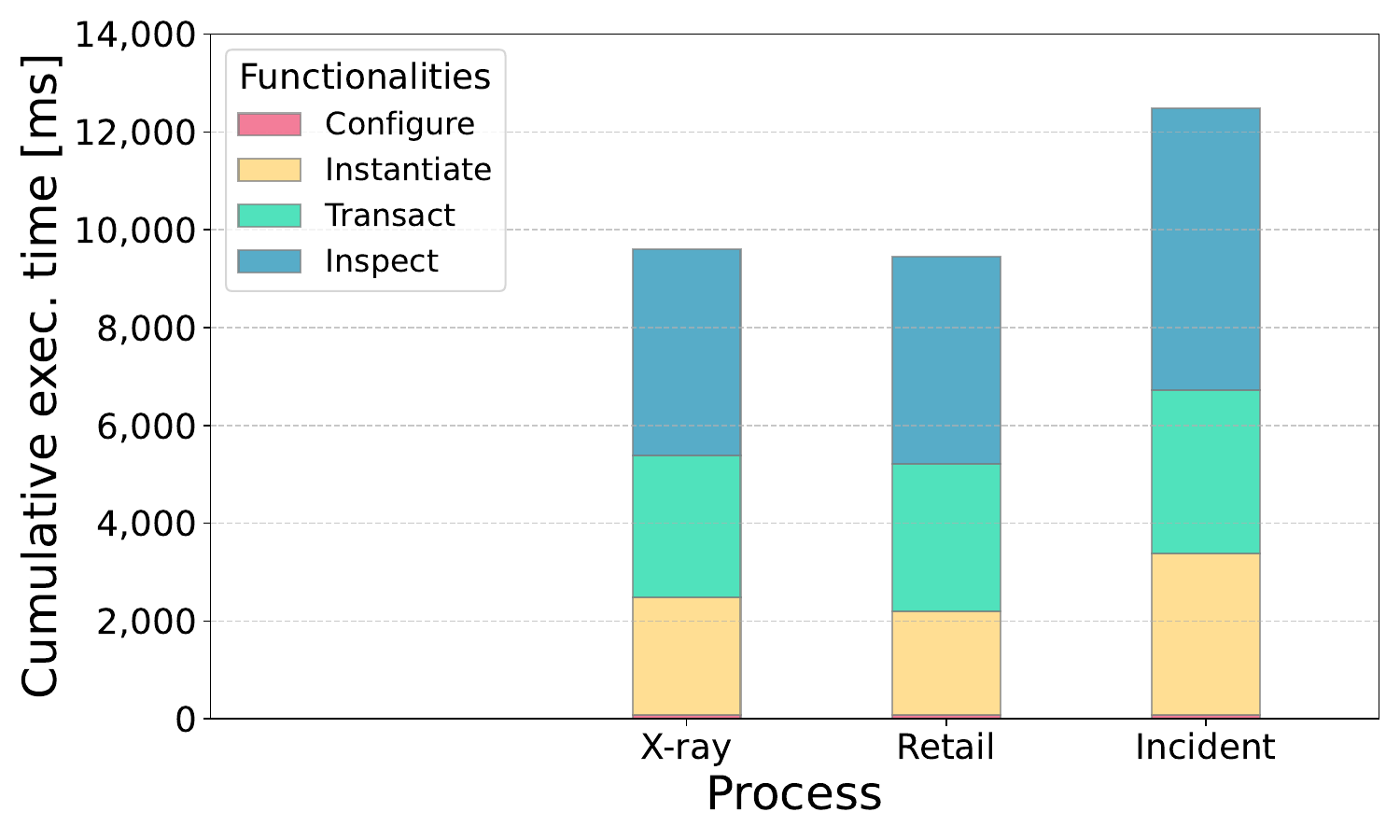} }  
		\label{fig:read_pub:perf}}
	\subfloat[Cost analysis]{{\includegraphics[width=0.5\textwidth]{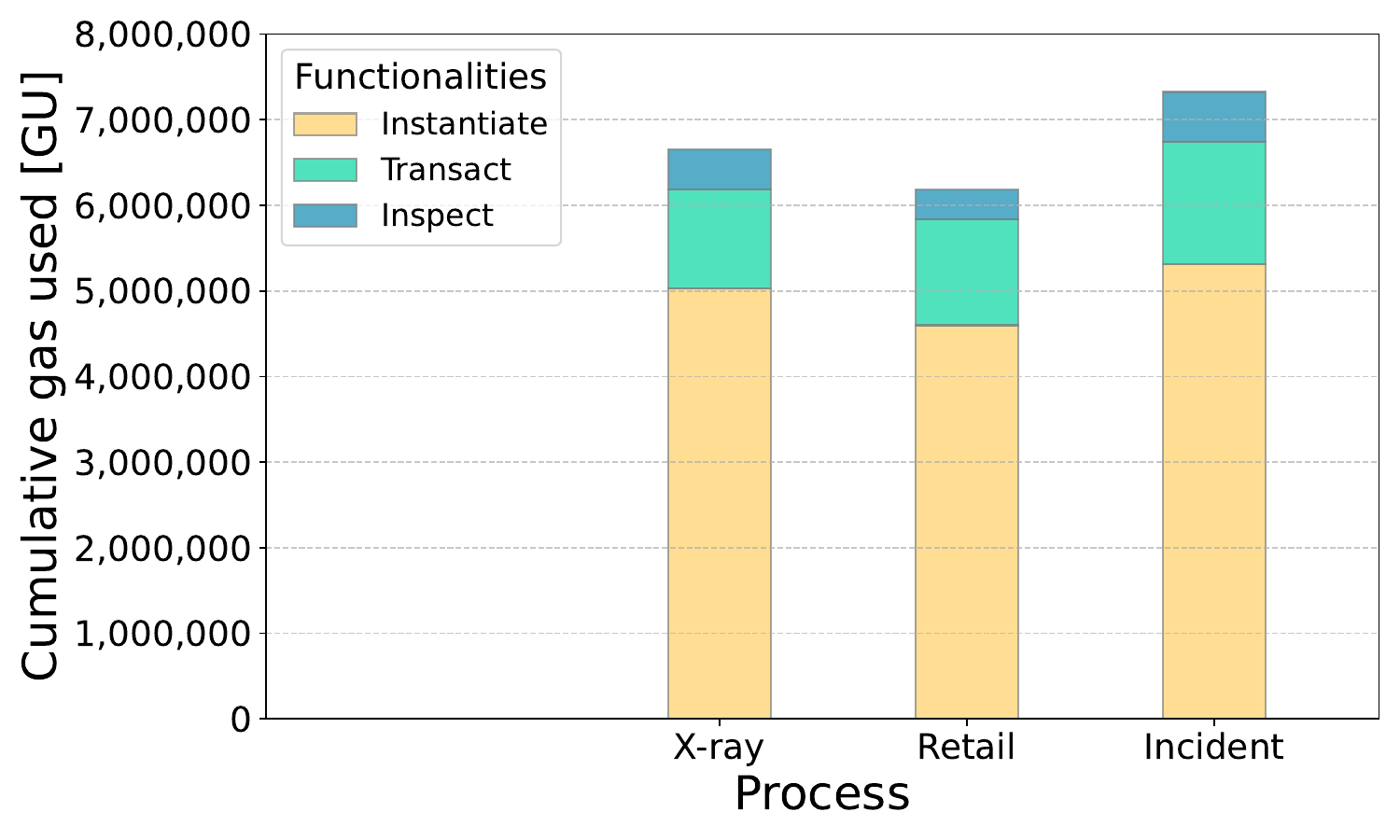} }
		\label{fig:read_conf:perf}}
	%\vspace{-1em}
	\caption{Performance comparison}
	\label{fig:comparison:analysis}
\end{figure}
\subsubsection{Execution Performance and Expenditure.}
We gauged the time and cost of our prototypical implementation against data simulating the execution and message exchange of processes from the literature. 
\Cref{fig:comparison:analysis} illustrates the results, averaging over runs that took the longest path.
We gathered data using a local RPC as a stub for the Programmable Blockchain Platform. 

Firstly, let us focus on simulations of our motivating scenario (\cref{sec:example}), which we will henceforth name X-ray process.
For the time evaluation, we ran an instance for the longest path five times.
We employed the Sepolia testnet as the Blockchain Platform and passed input data based on real-world files in the healthcare domain upon anonymization (available in our codebase). 
The \emph{configure} functionality turns out to be the fastest, requiring an average of \qty{71}{\ms}. % to retrieve and process the data and generate the policies.
For the \emph{instantiate} functionality, we let four process participants interact with the tool as specified in the choreography %. To accomplish this functionality, 
and it averaged \qty{2689}{\ms}. 
This higher amount depends on the need to authenticate process participants and prepare the data to send on-chain.
  
%The \emph{transact} functionality pertains to the exchange of choreography messages between participants according to the possible execution paths. These refer to the decision points on the \SmallCode{accepted} and \SmallCode{registration} variables. With the shortest path, an instance encompasses eight messages (with \SmallCode{accepted} set as \SmallCode{true} and \SmallCode{registration} \SmallCode{false}). The longest one includes a loop (with \SmallCode{accepted} set as \SmallCode{false} first, and \SmallCode{true} afterwards, and \SmallCode{registration} set to \SmallCode{false}) and entails the exchange of twelve messages. 
% pertains to the exchange of choreography messages between participants according to the followed execution path. In the considered experiments, we refer to the decision points on the \SmallCode{accepted} and \SmallCode{registration} variables. The 
%we considered the path that does not include a loop (with \SmallCode{accepted} and \SmallCode{registration} set as \SmallCode{false}) and entails the exchange of ten messages. 
For the \emph{transact} functionality, the runtime required an average of \qty{2476}{\ms} in total, with the encryption of single messages averaging \qty{247}{\ms}. 
%, ranging from \qty{4312}{\ms} to \qty{15739}{\ms}. 
%Finally, the \textit{inspect} functionality required from a minimum of \qty{7457}{\ms} to a maximum of \qty{25525}{\ms} with an average of \qty{18159}{\ms}.
Finally, the \textit{inspect} functionality required an average of \qty{4390}{\ms} %The higher time requested %for reading information  
including the key request management. However, notice that this step was performed only once %, at the beginning of the \textit{inspect}, 
and required \qty{2695}{\ms}. Once the key is received, the single reading and decryption operations for a message averaged \qty{169}{\ms}. 
Overall, if compared to the expected time for Ethereum block creation (around \qty{12}{\s}), the overhead introduced by \tool and related operations (e.g., encrypt and decrypt of message payloads) is orders of magnitude smaller (milliseconds) and can therefore be considered acceptable.
%Referring to the encryption and decryption of message payloads, the \tool overhead can be considered acceptable. 
%\Cref{tab:cost_performance} lists the units of gas required to run the X-ray process.  
To analyze the cost, we considered the sole operations that involve the blockchain, thus, \textit{instantiate}, \textit{transact}, and \textit{inspect}. %, plus the one-time bootstrapping cost of deploying smart contracts on-chain.
The highest cost is associated with the \textit{instantiate} functionality, consuming around \qty{5} million Gas Units (henceforth, GU), which corresponds to the total amount for the instantiation of the public state (\qty{2931698} GU) and other related operations (each averaging \qty{90577} GU). 
\textit{Transact}ing consumed \qty{1689190} GU in total for all the information exchanges, with a cost for a single transaction averaging at \qty{115096} GU.
Another relevant cost is the deployment of smart contracts, consuming around  \qty{5} million GU. However, this operation is performed only once during the bootstrapping of \tool and does not pertain to single process instantiations. 
%The highest costs are indeed associated with the deployment of smart contracts and the \textit{instantiate} functionality, consuming around \qty{5} million units of gas each. For the entire execution of the \textit{transact} functionality comprehending the ten messages, the total amount of gas units consumed is \qty{1689190}.
%, considering the average-length path (with both the \SmallCode{accepted} and \SmallCode{registration} variables set as \SmallCode{true}).
Finally, \textit{inspect} required % functionality appears to be the cheapest, with a total of 
\qty{89535} GU for the key request management. The reading operations do not produce transactions, thus incurring no fees. 
In total, the execution of the X-ray process, without considering the deployment, consumed a total of \qty{6884376} GU which translates to \qty{0.105035017}~ETH in fees.
However, to evaluate the actual cost in fiat currency, the fee value has to be related to a specific blockchain platform, as this aspect significantly influences the multiplicative factor. Considering Polygon blockchain, the above fee of \qty{0.105035017}{POL} would result in \qty{0.02}{US\$}\footnote{The conversion rate is of \qty{0.21}{US\$} at the time of writing (March 14, 2025)},
representing an affordable cost considering the obtained benefits. 
%Compared to 
Taking as a reference another blockchain-backed entity like a standard Ethereum Non-Fungible Token (NFT), it requires around \qty{2} million GU for deployment and \qty{50000} to \qty{200000} {GU} for transfer or minting operations~\cite{choi2023gas}. In this context, \tool exhibits a higher deployment cost (performed only once for kickstarting the entire \tool system) given by the more complex smart contracts logic, while its execution cost is in line with an NFT, resulting in a cost-efficient performance. 

%\Cref{fig:comparison:analysis} shows the execution time and costs registered during the execution of the X-ray diagnostic process. For the time evaluation, we ran a separate process instance for each path in the control flow (each corresponding to a different decision in correspondence with the exclusive-choice gateways) five times (thus totalling twenty executions). \cref{tab:time_performance}. We employed the Sepolia testnet as the Blockchain Platform and passed input data based on real-world files in the healthcare domain upon anonymization. The dataset and the Sepolia transaction handles are openly accessible at \href{anonymous.4open.science/r/CONFETTY/}{\nolinkurl{github.com/apwbs/CONFETTY}}.
% \subsubsection{Comparison with the literature.}
%the time and cost needed by \tool to run three 
Let us now compare the above results
with two other processes from the literature, namely 
%the aforementioned X-ray~\cite{runningExample}, 
retail~\cite{Corradini.etal/BCRA2021:ModelDrivenEngineering} and incident management~\cite{Corradini.etal/ACMTMIS2022:EngineeringChoreographyBlockchain}. 
%different branches in correspondence with exclusive choice gateways. 
%
%   X-ray analysis	- incident    - retail
%       10 msg      -  11 msg     - 12 msg
%      5 gateways	- 3 gateways  - 2 gateways
%      4 parties    - 5 parties   - 3 parties
%The three processes 
The  X-ray, incident and retail processes have similar structures in the number of messages (10, 13 and 12, respectively), gateways (5, 3 and 2) and involved participants (4, 5 and 3) and exhibit similar trends from both runtime and cost standpoints. 
The incident management process is the slowest and most expensive one %. However, it shows only 
with an \SIlist{8;10}{\percent} increase in cumulative time and %a \SI{10}{\percent} increase in cumulative 
cost compared to the X-ray process, respectively.
%This increase is influenced by the number of participants, process elements, and exchanged messages, which result in additional steps and transactions.
% (\num{5} in the incident management against the \num{3} of the X-ray process), thus leading to the execution of more steps.
The experiments show that \tool performs comparably to literature tools and scenarios. 
%like to various processes, enabling the comparison of its results with the tools that execute the processes taken from the literature. 
Indeed,
ChorChain \cite{Corradini.etal/ACMTMIS2022:EngineeringChoreographyBlockchain} required approximately \num{5} million GU for the incident management process, while MultiChain \cite{Corradini.etal/BCRA2021:ModelDrivenEngineering} required around \num{6} million GU for the retail process.
Notice that such approaches only focus on public execution and enforcement without incorporating any confidential mechanisms and that they deploy a contract for each new instance.
%
%Overall, the comparison demonstrates that \tool performance is in line with the literature. Furthermore, differently from the mentioned solutions, \tool deploys a single contract for the entire architecture rather than a separate contract for each instance, leading to a more cost-efficient execution. 

\subsubsection{Scalability Assessment.}
\begin{figure}[tb]
	\subfloat[Time analysis]{{\includegraphics[width=0.5\textwidth]{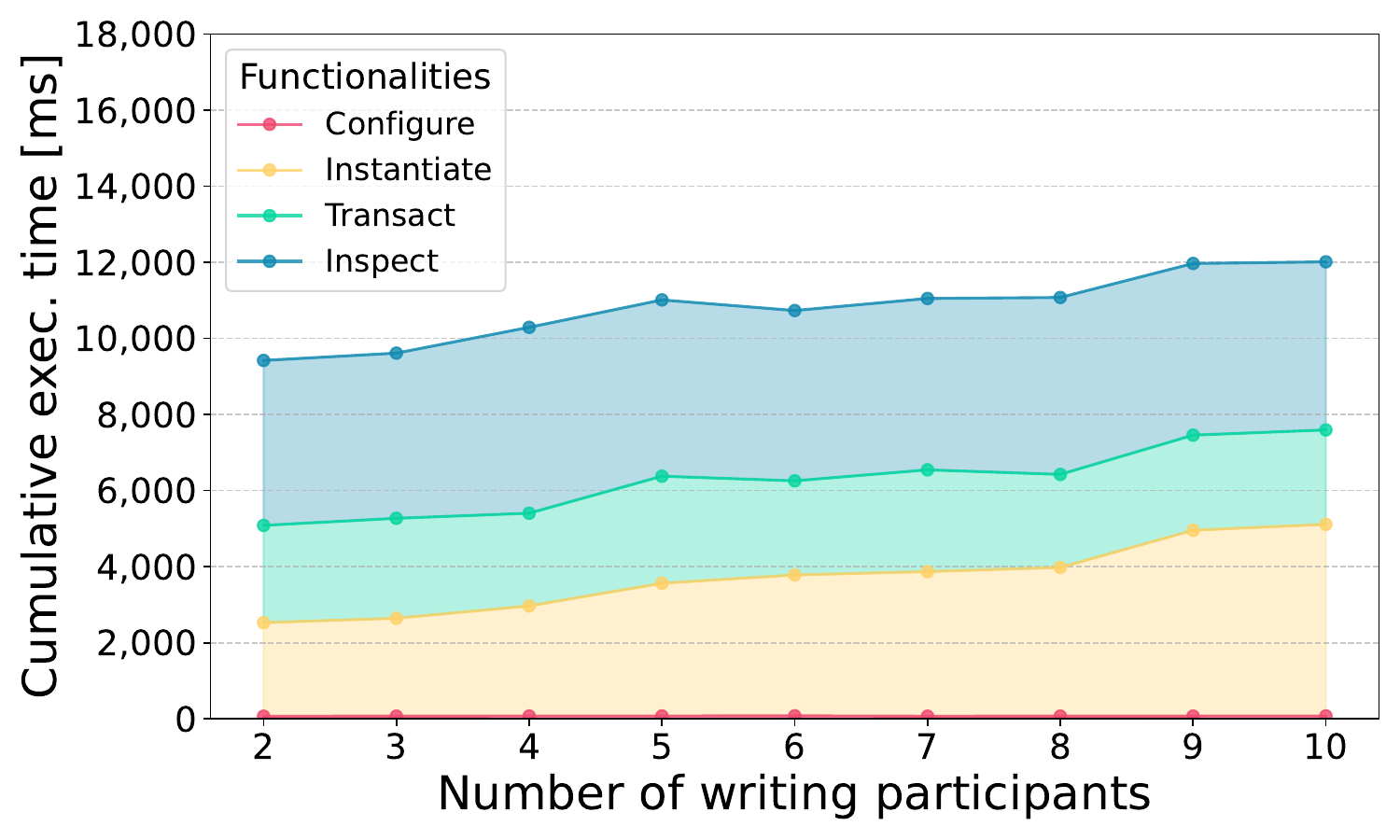} }  
		\label{fig:time:encrypters}}
	\subfloat[Cost analysis]{{\includegraphics[width=0.5\textwidth]{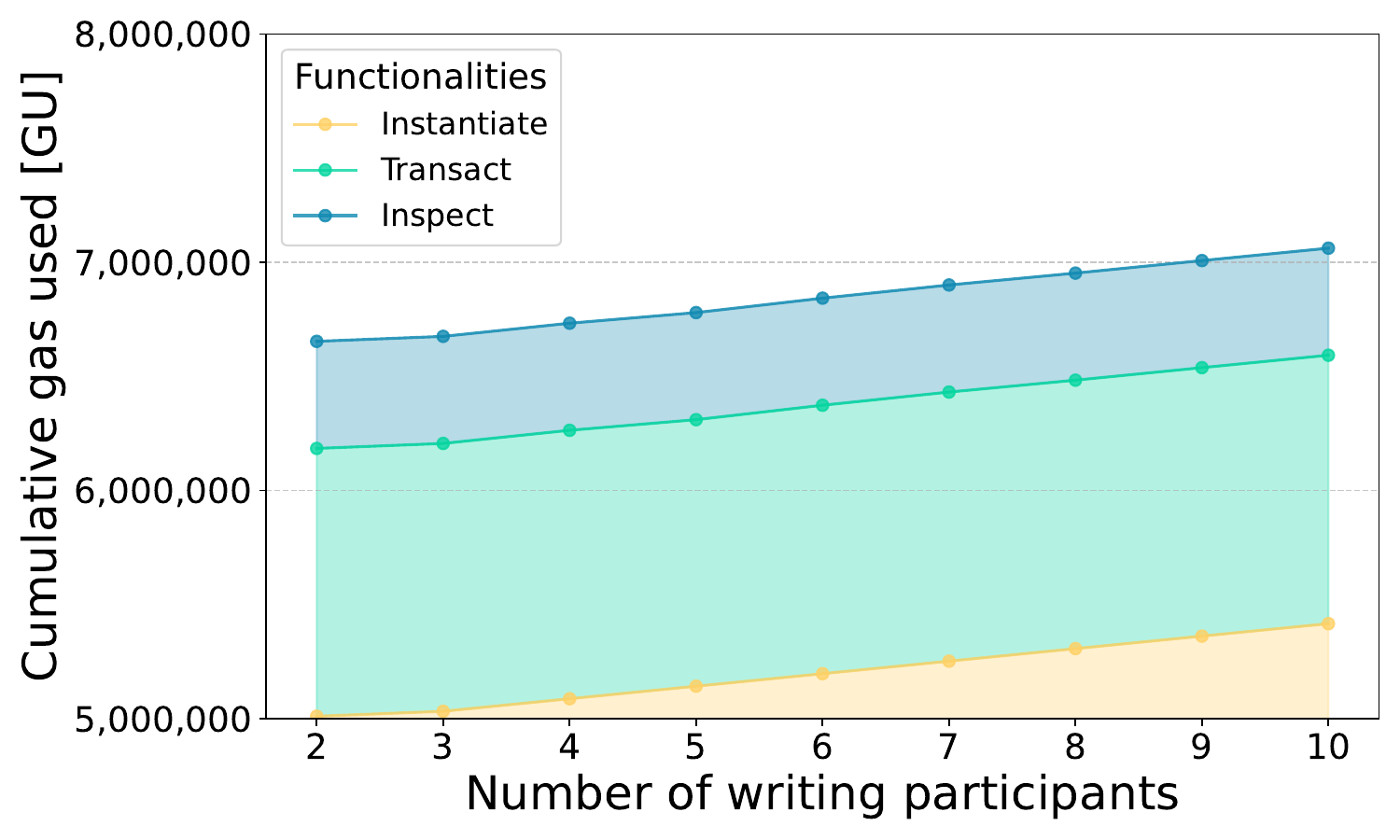} }
		\label{fig:gas:encrypters}}
	%\vspace{-1em}
	\caption{Performance with increasing process participants}
	\label{fig:comparison:encrypters}
\end{figure}

\begin{figure}[tb]
	\subfloat[Time analysis]{{\includegraphics[width=0.5\textwidth]{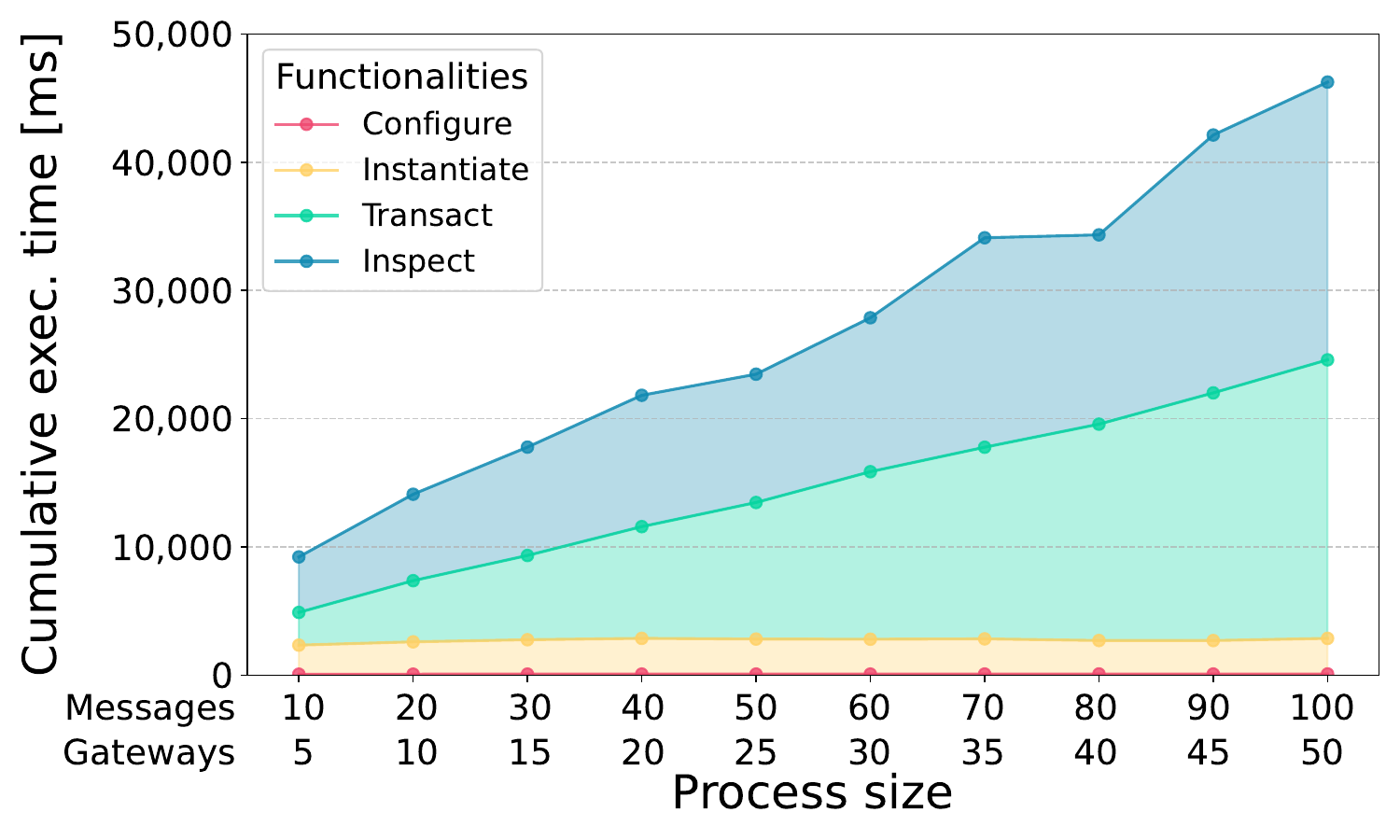} }  
		\label{fig:time:loop}}
	\subfloat[Cost analysis]{{\includegraphics[width=0.5\textwidth]{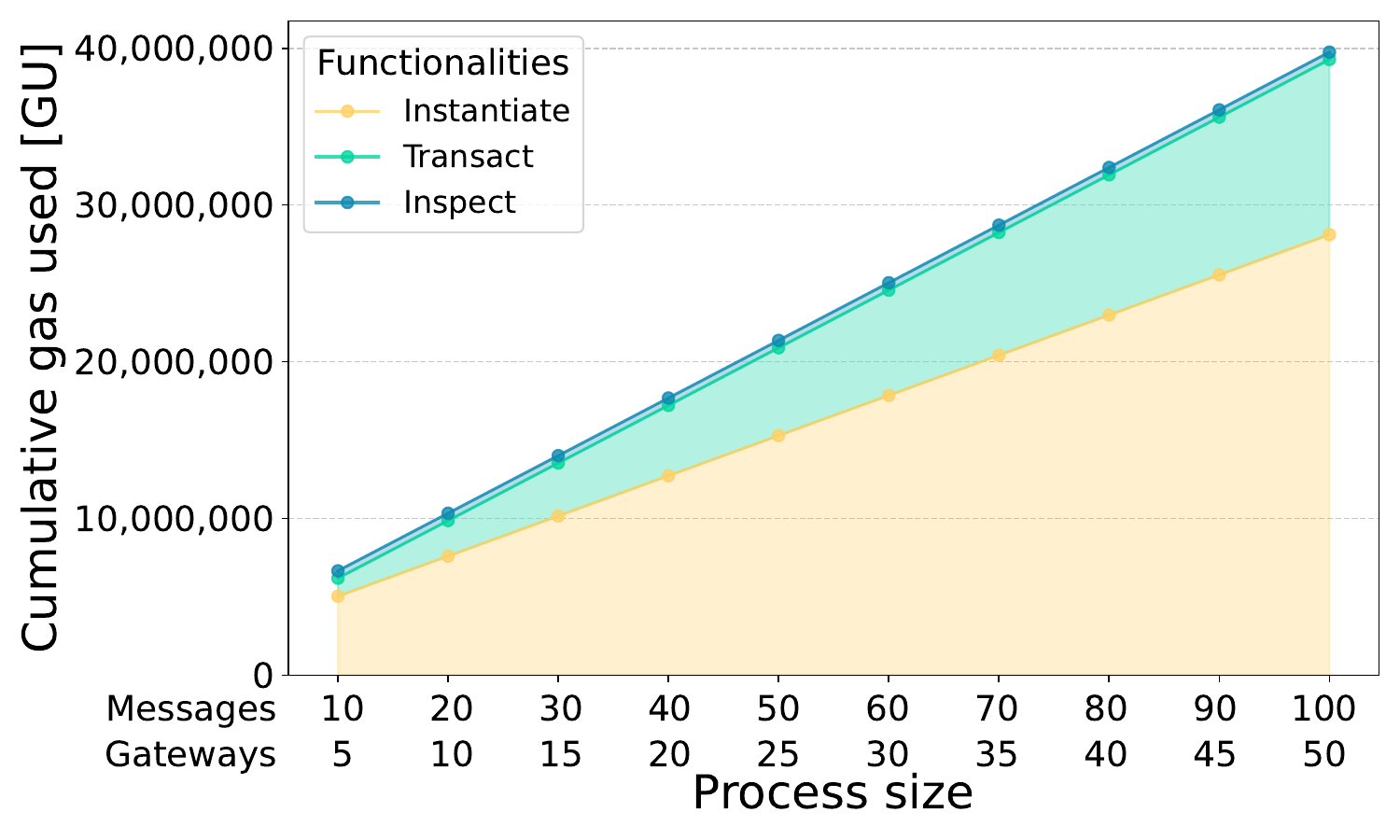} }
		\label{fig:gas:loop}}
	%\vspace{-1em}
	\caption[Performance with increasing choreography size]{Performance with increasing choreography size}
	\label{fig:comparison:loop}
\end{figure}

Lastly, we analyzed the scalability of \tool.
To this end, we considered the dependence of time and cost taking as independent variables the dimensions of the process affecting the input size for our tool. In this section, we focus on the number of process participants and the choreography model size due to space restrictions as they evidence the most salient results.
% we report only noteworthy experiments and outcomes 
%while we mention a highlight of the others. 
%while the full 
Further experiments including other variables like the number of gateways (parallel and exclusive) and the message payload size can be found in the online repository. 
The default values for the variables adhere to the example provided in \cref{fig:running_example}. For every test, we let one variable change the value and keep the others fixed, assigned with the respective defaults.
%at \href{anonymous.4open.science/r/CONFETTY/}{\nolinkurl{anonymous.4open.science/r/CONFETTY/}}.    
\Cref{fig:comparison:encrypters,fig:comparison:loop} dissect the registered timings and costs to serve the different functionalities concurring to the total by stacking the respective subtotals.

\Cref{fig:comparison:encrypters} depicts the performance trend % first changed variable, the
varying the number of process actors actively sending messages %. The execution included an increasing number of participants, ranging 
from two (the minimum for a multi-party process), to ten (the maximum admissible within the X-ray choreography, with one different sender per message and activity). %All the other variables of the process were kept fixed.
\Cref{fig:time:encrypters} depicts the trend for cumulative time performance, showing a total variation of \qty{3.9}{\s}, mostly related to the \textit{instantiate} functionality as it includes the authentication of all the participants. %, hence being the most influenced by the test. 
The costs shown in \cref{fig:gas:encrypters} follow a comparable linear trend. %shows the registered costs, following a comparable trend. 
%are influenced in the same manner and exhibit a linear growth similar to the previous case. 
%instead that no relevant behaviors occur on the blockchain as the number of participants is related to off-chain operations.
%and the size of the exchanged messages. For these two cases, we ran ten different instances of the choreography, keeping stable the number of elements and the executed path while changing the other factors. In the first case, the execution included an increasing number of process participants, ranging from two, the minimum case for a multi-party process, to ten, the maximum number admissible within the considered choreography (one per message). In the second case, the execution involved increasing the size of message payloads, starting from the base case and scaling up to ten times the original size.

\Cref{fig:comparison:loop} 

displays the results observed varying the process model size. % To assess the scalability, w
We progressively increased it %the choreography size 
by sequentially concatenating the whole choreography diagram multiple times to derive a new one, thus scaling the size up to ten times the original one. %As a result, we increased the number of elements (\eg messages and gateways) and kept the number of participants and the message payload size fixed. 
\Cref{fig:time:loop}
shows a linear growth in the runtime of the \textit{transact} and \textit{inspect} functionalities, without a significant impact on the remaining one, \textit{instantiate}. 
The execution time increases with the number of messages due to additional read and write operations. 
%This behavior depends on the number of messages, as the incremental number leads to more read and write operations, affecting the total execution time.
\Cref{fig:gas:loop} shows the cost increase, with a linear growth registered for the \textit{instantiate} and \textit{transact} functionalities. More elements increase on-chain storage needs and hence gas cost. Read operations remain constant as they only handle one-time key management on-chain (hence the flat cost associated with \textit{inspect}). 

\section{Conclusions}
\label{sec:conclusions}
In this work, we presented \tool, a blockchain-based architecture for PAISs. \tool enables transparency and public enforcement for process execution while preserving the confidentiality of sensitive exchanged data.
A public state saved on-chain maintains process data for enforcement and auditability purposes. The evolution of the public state is mediated by smart contracts implementing the business process control- and data-flow logic. We use \MAABE encryption to hide sensitive information and grant access only to authorized parties based on their roles, encoded as attributes in access policies and embedded directly in their decryption keys. %To evaluate the \tool architecture, w
We confirmed its security guarantees with a threat model analysis, and demonstrated the time and cost performance of our implemented prototype with processes taken from the literature. 

In future work, we envision a field study to assess the practical adoption and usability with real-world stakeholders. Another interesting avenue is the full decentralization of the architecture (to remove the existing points of trust residing in off-chain components) and governance (to cater for multiple process owners during the instantiation phase). Conducting a security analysis of our approach against stronger adversarial models is an intriguing future endeavor, too. Finally, we envision a study and design of a solution to enable automated decision support via computation held on encrypted data. 

\noindent\parbox{\textwidth}{
	\subsubsection{\ackname}
	This work was partly funded by projects PINPOINT (B87G22000450001) under the PRIN MUR program, and Health-e-Data, funded by the EU-NGEU under the Cyber~4.0 NRRP MIMIT programme.
}

%
% ---- Bibliography ----
%
% BibTeX users should specify bibliography style 'splncs04'.
% References will then be sorted and formatted in the correct style.
%
\bibliographystyle{splncs04}
\bibliography{biblio}
\end{document}